\documentclass{article}
\usepackage{amssymb}
\usepackage{amsmath}

\input{tcilatex}
\begin{document}

\begin{center}
\textbf{REVISITING PERFECT FLUID DARK MATTER:\ OBSERVATIONAL CONSTRAINTS
FROM OUR GALAXY}

\bigskip

\bigskip

Alexander A. Potapov$^{1,a}$, Guzel M. Garipova$^{1,b}$ and Kamal K. Nandi$%
^{1,2,3,c}$

$\bigskip $

$^{1}$Department of Physics \& Astronomy, Bashkir State University,
Sterlitamak Campus, Sterlitamak 453103, RB, Russia

$^{2}$Zel'dovich International Center for Astrophysics, M. Akmullah Bashkir
State Pedagogical University, Ufa 450000, RB, Russia

$^{3}$Department of Mathematics, North Bengal University, Siliguri 734 013,
WB, India

$\bigskip $

$^{a}$E-mail: potapovaa@mail.ru

$^{b}$E-mail: goldberg144@gmail.com

$^{c}$E-mail: kamalnandi1952@yahoo.co.in
\end{center}

PACS numbers: 04.20 Gz,04.50 + h, 04.20 J

\begin{center}
\bigskip -----------------------------------------------------

\bigskip

\textbf{Abstract}

\bigskip
\end{center}

We revisit certain features of an assumed spherically symmetric perfect
fluid dark matter halo in the light of the observed data of our galaxy, the
Milky Way (MW). The idea is to apply the Faber-Visser approach of combined
observations of rotation curves and lensing to a first post-Newtonian
approximation to "measure" the equation of state $\omega (r)$ of the perfect
fluid galactic halo. However, for the model considered here, no constraints
from lensing are used as it will be sufficient to consider only the rotation
curve observations. The lensing mass together with other masses will be just
computed using recent data. Since the halo has attractive gravity, we shall
impose the constraint that $\omega (r)\geq 0$ for $r\leq R_{\text{MW}}$,
where $R_{\text{MW}}\sim 200$ kpc is the adopted halo radius of our galaxy.
The observed circular velocity $\ell $ $\left( =2v_{\text{c}%
}^{2}/c_{0}^{2}\right) $ from the flat rotation curve and a crucial
adjustable parameter $D$ appearing in the perfect fluid solution then yield
different numerical ranges of $\omega (r)$. It is demonstrated that the
computed observables such as the rotation curve mass, the lens mass, the
post-Newtonian mass of our galaxy compare well with the recent mass data. We
also calculate the Faber-Visser $\chi -$ factor, which is a measure of
pressure content in the dark matter. Our analysis indicates that a range $%
0\leq \omega (r)\leq 2.8\times 10^{-7}$ for the perfect fluid dark matter
can reasonably describe the attractive galactic halo. This is a strong
constraint indicating a dust-like CDM halo $(\omega \sim 0)$ supported also
by CMB constraints.

\bigskip

\textbf{Keywords:}\textit{\ Dark matter, perfect fluid, equation of state,
galactic masses}

\section{Introduction}

A few years ago, in Ref.[1], a perfect fluid dark matter model was developed
that was shown to have many attractive theoretical aspects. The solution may
be thought of as a dark matter induced spacetime embedded in a static
cosmological Friedmann-Lema\^{\i}tre-Robertson-Walker (FLRW) background%
\footnote{%
The reason for the appearance of static FLRW background around the imbedded
perfect fluid dark matter is already explained in Ref.[1]. The Einstein
field equations are solved with perfect fluid stress tensor in both the
cases but we sought a static solution from the start. While working on a
local problem (flat rotation curve), the scale factor is usually fixed to $%
R_{0}=1$ today.}. The motivation for developing an isotropic perfect fluid
model (we leave open the question of particle identity of dark matter) came
from the fact that predictions from such model at stellar and cosmic scales
have been observationally well corroborated so far. More recently, Harko and
Lobo [2] \ investigated dark matter as a mixture of two non-interacting
perfect fluids, with different four-velocities and thermodynamic parameters.
Gonz\'{a}lez-Morales and Nu\~{n}ez [3] compared two different dark matter
models, one is a perfect fluid and the other is a scalar field [3]. See also
[4].

The model considered here assumes that a spherical dark matter distribution
is the only gravitating source. This assumption is of course an
oversimplification since, although the bulge is quite spherical and is
dominated by old stars, the Milky Way has a strongly flattened stellar
distribution. However, we know from the vertical velocity dispersion of
stars as a function of distance from the disk plane that the local disk mass
density is almost identical to the sum of the densities that can be
attributed to stars, gas and stellar remnants. Therefore, there is
practically little dark matter hidden in the disk. Hence, to explain the
rotation curve measurements, we are forced to assume that dark matter
resides in the halo region dominating its mass, \textit{is} spherically
distributed and, if it is non-baryonic, would not be expected to collapse
into a disk-like structure.

Specifically, the hypothesis of dark matter arose because the Newtonian
circular velocity $v_{\text{c}}^{2}=\frac{GM(r)}{r}$ of circularly moving
probe particles caused by the luminous mass distribution $M(r)$ is not
supported by observations [5,6]. The circular velocity becomes nearly flat, $%
v_{\text{c}}^{2}\simeq $ constant, at distances far away from the center
(halo region), which is possible only if $M(r)\propto r$. Therefore, almost
every galaxy is assumed to host a large amount of non-luminous matter, the
so called gravitational dark matter, consisting of unknown particles not
included in the particle standard model, forming a halo around the galaxy.
Naturally, dark matter is at the core of modern astrophysics. Many well
known theoretical models for dark matter exist in the literature, for
instance, see [7-29] (the list is by no means exhaustive). Some models that
do not hypothesize dark matter appear in [30-38]. Well known density
profiles originated in [39-41]. Excellent reviews are to be found in [42-45].

In this paper, we shall revisit the model of perfect fluid dark matter,
developed in Ref.[1], in the light of the observed/inferred data of our
galaxy. Our analysis would require three ingredients: (i) A method, viz.,
the Faber-Visser [46] method of combined post-Newtonian measurements of
rotation curves and gravitational lensing for measuring the equation of
state $\omega (r)$ of the dark matter and determining the rotation curve
mass ($m_{\text{RC}}$), the lens mass ($m_{\text{Lens}}$) and the
post-Newtonian mass ($M_{\text{pN}}$). However, for the perfect fluid
solution we consider here, it suffices to consider only the rotation curve
as a constraint, while the lens mass will be a result of computation. (ii)
Two observed inputs, viz., the circular velocity $\ell $ $\left( =2v_{\text{c%
}}^{2}/c_{0}^{2}\right) $ of probe particles, where $c_{0}$ is the speed of
light in vacuum, and the radius $R_{\text{MW}}$ of our galactic halo. (iii)
An observational constraint, viz., the one imposed by the attractive nature
of dark matter so that $p/\rho =\omega (r)\geq 0$. The nature is attractive
because the very existence of dark matter is speculated from observations of
the Doppler shifted light emanating from neutral hydrogen clouds moving on
stable circular orbits in the galactic halo [20,32,47,48]. Using these
ingredients, we shall analyze how choices of the adjustable parameter $D$
appearing in the dark matter metric lead to different types of scenarios.

The following are our new results: Depending on the values of $D$, we show
that (i) The observable masses $m_{\text{RC}}$, $m_{\text{Lens}}$ and $M_{%
\text{pN}}$ compare well with the masses inferred by other independent
means. (ii) There could appear an intriguing negative pressure matter sector
($\omega <0$) beyond the halo radius\footnote{%
However, it will be shown later that the $\omega <0$ matter sector is not
exotic in nature. It will also be evident that, we can shift the values of $%
D $ to make $\omega <0$ matter appear at any finite radius beyond halo
radius $R_{\text{MW}}$, but we must take care not to violate the attractive
nature $\omega \geq 0$ \textit{inside} $R_{\text{MW}}.$}. (iii) The
Faber-Visser $\chi -$ factor has values near unity so that pressure
contribution to the post-Newtonian mass $M_{\text{pN}}$ is negligible. Hence
the perfect fluid dark matter resembles dust ($\omega \sim 0$) akin to CDM
model. (iv) There is flexibility in the halo radius in the sense that our
model can accommodate extended radii. All these imply that, fundamentally,
the perfect fluid model stands up to actual observations on mass, equation
of state and in addition predicts marginal quitessence matter at asymptotic
distances, all within a single formulation.

In Sec.2. we briefly outline the perfect fluid dark matter and in Sec.3,
display the Faber-Visser post-Newtonian observables. In Sec.4, we apply the
galaxy inputs to those observables and deduce the most suitable range of $D$
that agrees with the observational constraints from our galaxy. In Sec.5, we
conclude the paper. We take $G=1,$ $c_{0}=1$, unless specifically mentioned.

\section{Perfect fluid dark matter}

\bigskip The general static spherically symmetric space-time is represented
by the following metric

\begin{equation}
ds^{2}=-e^{\nu (r)}dt^{2}+e^{\lambda (r)}dr^{2}+r^{2}(d\theta ^{2}+\sin
^{2}\theta d\varphi ^{2})\text{ ,}
\end{equation}%
where the functions $\nu (r)$ and $\lambda (r)$ are the metric potentials.
For the perfect fluid, the matter energy momentum tensor $T_{\beta }^{\alpha
}$ is given by $T_{t}^{t}=\rho (r)$, $T_{r}^{r}=T_{\theta }^{\theta
}=T_{\varphi }^{\varphi }=p(r)$, where $\rho (r)$ is the energy density, $%
p(r)$ is the isotropic pressure. Considering flat rotation curve as an
input, an exact solution of Einstein field equations is derived in [1]:

\begin{equation}
e^{\nu (r)}=B_{0}r^{\ell }\text{,}
\end{equation}

\begin{equation}
e^{-\lambda (r)}=\frac{c}{a}+\frac{D}{r^{a}},
\end{equation}%
\begin{equation}
a=-\frac{4(1+\ell )-\ell ^{2}}{2+\ell },
\end{equation}

\begin{equation}
c=-\frac{4}{2+\ell }\text{ ,}
\end{equation}

\begin{equation}
\ell =2v_{\text{c}}^{2}/c_{0}^{2},
\end{equation}%
where $B_{0}>0$, $D$ are integration constants and $v_{\text{c}}$ is the
circular velocity of stable circular hydrogen gas orbits treated as proble
particles. The exact energy density and pressure are%
\begin{eqnarray}
\rho (r) &=&\frac{1}{8\pi }\left[ \frac{\ell (4-\ell )}{4+4\ell -\ell ^{2}}%
r^{-2}-\frac{D(6-\ell )(1+\ell )}{2+\ell }r^{\frac{\ell (2-\ell )}{2+\ell }}%
\right] \\
p(r) &=&\frac{1}{8\pi }\left[ \frac{\ell ^{2}}{4+4\ell -\ell ^{2}}%
r^{-2}+D(1+\ell )r^{\frac{\ell (2-\ell )}{2+\ell }}\right] .
\end{eqnarray}

The free adjustable parameter $D$, having the dimension of (length)$^{-2}$,
in the solution is extremely sensitive and its value can be decided only by
observed physical constraints. In the present case, the constraint is that
the galactic fluid be non-exotic and attractive, i.e., the equation of state
parameter $\omega (r)=\frac{p(r)}{\rho (r)}\geq 0$ must hold within the halo
radius. With this information at hand, an interesting aspect of the solution
can be found from Eqs.(7) and (8).

It can be seen that the integrated quantity, call it $M_{0}=4\pi
\dint\limits_{0}^{r}\rho (r)r^{2}dr$ derived from exact $\rho (r)$ given by
Eq.(7), is identical with the Newtonian mass $M_{\text{N}}$ derived in
Eq.(23) below. One could as well call $M_{\text{N}}$ the post-Newtonian
counterpart of $M_{0}$ since $\rho (r)$ in Eq.(23) is expressed as
derivatives of post-Newtonian masses. The question then we ask: What
quantity derived from the exact solution differs from its measurable
post-Newtonian counterpart? One such quantity is the total mass within a
radius $r$ with \textit{pressure contribution}, which is defined by, using
Eqs.(7) and (8)%
\begin{eqnarray}
M_{\text{total}}(r) &=&4\pi \dint\limits_{0}^{r}(\rho +3p)r^{2}dr=\frac{\ell
(2+\ell )r}{4+4\ell -\ell ^{2}}+\frac{2D}{\ell -6}r^{\frac{4+4\ell -\ell ^{2}%
}{2+\ell }} \\
&=&\frac{\ell r}{2}+\frac{D\ell r^{3}}{3}-\frac{\ell ^{2}r}{4}+O(\ell ^{2}).
\end{eqnarray}%
We shall see in the next section that its post-Newtonian counterpart is just 
$M_{\text{pN}}(r)=\frac{\ell r}{2}$. Hence the theoretical and observable
masses in principle differ depending on arbitrary values of $D$, even when $%
D=0$. Therefore, let us proceed to define the post-Newtonian observables.

\section{Faber-Visser post-Newtonian observables}

We shall only quote the relevant expressions here. For details, see
Faber-Visser [46]. They considered the metric in the form

\begin{equation}
ds^{2}=-e^{2\Phi (r)}dt^{2}+\frac{dr^{2}}{1-\frac{2m(r)}{r}}+r^{2}(d\theta
^{2}+\sin ^{2}\theta d\varphi ^{2})\text{ ,}
\end{equation}%
which is completely determined by the two metric functions $\Phi (r)$ and $%
m(r)$. Comparing it with the metric (1), we have

\begin{equation}
m(r)=\frac{r(1-\frac{c}{a}-\frac{D}{r^{a}})}{2}\text{,}
\end{equation}

\begin{equation}
\Phi (r)=\frac{\log (B_{0})\text{ }+\text{ }\ell \log (r)\text{ }}{2}.
\end{equation}%
The potentials $\Phi _{\text{RC}}(r)$ and $\Phi _{\text{Lens}}(r)$, obtained
respectively from the rotation curve data and gravitational lensing
observations, are derived to be%
\begin{equation}
\Phi _{\text{RC}}(r)=\Phi (r)=\frac{\log (B_{0})\text{ }+\text{ }\ell \log
(r)\text{ }}{2}\text{,}
\end{equation}

\begin{eqnarray}
\Phi _{\text{Lens}}(r) &=&\frac{\Phi (r)}{2}+\frac{1}{2}\dint \frac{m(r)}{%
r^{2}}dr=\frac{\log (B_{0})\text{ }+\text{ }\ell \log (r)\text{ }}{4}  \notag
\\
&&+\frac{D(\ell +2)r^{\frac{4(1+\ell )-\ell ^{2}}{2+\ell }}+\ell (\ell
-4)\log (r)}{4(\ell ^{2}-4\ell -4)}\text{ .}
\end{eqnarray}%
The lensing potential $\Phi _{\text{Lens}}(r)$ is a fundamental observable
quantity. When the pressures and matter fluxes are small compared to the
mass-energy density then $\Phi _{\text{RC}}(r)=\ \Phi _{\text{Lens}}(r)$,
otherwise they may not be equal.

One pseudo-mass, inferred from rotation curve measurements, is given by

\begin{equation}
m_{\text{RC}}(r)=r^{2}\Phi ^{\prime }(r)=\frac{\ell r}{2}\text{.}
\end{equation}%
Another pseudo-mass $m_{\text{Lens}}(r)$, obtained from lensing
measurements, is defined as$\ $

\begin{equation}
\ m_{\text{Lens}}(r)=\frac{r^{2}\Phi _{\text{RC}}(r)}{2}+\frac{m(r)}{2}=%
\frac{r[a(1+\ell -Dr^{-a})-c]}{4a}\text{ .}
\end{equation}%
For the equation of state parameter for perfect fluid, we should evaluate $%
\omega $ and impose the constraint that up to $r=R_{\text{MW}}$,%
\begin{equation}
\omega (r)=\frac{p_{r}(r)+2p_{t}(r)}{\rho (r)}\geq 0,
\end{equation}%
which will provide a limit on $D$. From the first order approximations of
Einstein's equations, one obtains [46]

\begin{equation}
\rho (r)=\frac{2m_{\text{Lens}}^{\prime }(r)-m_{\text{RC}}^{\prime }(r)}{%
4\pi r^{2}}=\frac{r^{(-2-a)}\left[ -cr^{a}+a(r^{a}-D)+a^{2}D\right] }{8\pi a}%
\text{,}
\end{equation}

\begin{eqnarray}
p_{r}(r)+2p_{t}(r) &=&\frac{2\left[ m_{\text{RC}}^{\prime }(r)\ -m_{\text{%
Lens}}^{\prime }(r)\right] }{4\pi r^{2}}  \notag \\
&=&\frac{r^{(-2-a)}\left[ cr^{a}-a^{2}D+a\left\{ (\ell -1)r^{a}+D\right\} %
\right] }{8\pi a}\text{ .}
\end{eqnarray}%
Then Eq.(18) yields%
\begin{eqnarray}
\omega (r) &=&\frac{p_{r}(r)+2p_{t}(r)}{3\rho (r)}\approx \frac{2}{3}\frac{%
m_{\text{RC}}^{\prime }(r)\ -m_{\text{Lens}}^{\prime }(r)}{2m_{\text{Lens}%
}^{\prime }(r)-m_{\text{RC}}^{\prime }(r)}  \notag \\
&=&\frac{cr^{a}-a^{2}D+a\left[ (\ell -1)r^{a}+D\right] }{3\left[
-cr^{a}+a(r^{a}-D)+a^{2}D\right] }\text{ .}
\end{eqnarray}%
We have intentionally kept in the left hand side of the above Eq.(21) the
transverse pressure component $p_{t}$ for transparency, remembering that for
perfect fluid $p_{r}=p_{t}$, an exact equality that was used to derive the
metric (1).

It is to be emphasized that, \textit{observationally}, such exact equalities
as $p_{r}=p_{t}$ are impossible to attain. It follows that the difference in
dimensionless pressures is not zero but [46]%
\begin{eqnarray}
4\pi r^{2}\left[ p_{r}(r)-p_{t}(r)\right] &=&\frac{2}{r}\left( m_{\text{RC}%
}-m_{\text{Lens}}\right) -r\left[ \frac{m_{\text{RC}}-m_{\text{Lens}}}{r}%
\right] ^{\prime }+O\left( \frac{2m}{r}\right) ^{2}  \notag \\
&=&\frac{r^{-a}\left[ cr^{a}+2a^{2}D+a\left\{ (\ell -1)r^{a}+D\right\} %
\right] }{8\pi a},
\end{eqnarray}%
which is just the \textit{post-Newtonian version of isotropicity} of the
perfect fluid. However, this value of the right hand side for our galaxy
(and presumably for all other samples as well) is exceedingly small but not
exactly zero!

The next issue is whether the model is Newonian or not, that is, how much of
pressure contribution to mass is there. For this, we need to compare the two
integrals, one is the Newtonian mass $M_{\text{N}}(r)$ given by, using
Eqs.(19) and (20),

\begin{equation}
M_{\text{N}}(r)=4\pi \dint\limits_{0}^{r}\rho (r)r^{2}dr=\frac{%
r(a-c-ar^{-a}D)}{2a}\text{,}
\end{equation}%
and the other is the mass in the first post-Newtonian approximation [46]

\begin{equation}
M_{\text{pN}}(r)=4\pi \dint\limits_{0}^{r}(\rho +p_{r}+2p_{t})r^{2}dr=\frac{%
\ell r}{2}\text{.}
\end{equation}%
Eqs.(14-24) are the needed Faber-Visser post-Newtonian observables to be
examined using the available galactic data.

The Faber-Visser $\chi -$ factor, designed to provide a measure of the size
of the pressure contribution, can be obtained from Eq.(21)%
\begin{equation}
\chi (r)=\frac{m_{\text{Lens}}^{\prime }(r)}{m_{\text{RC}}^{\prime }(r)}=%
\frac{2+3\omega (r)}{2+6\omega (r)}.
\end{equation}%
For dust matter, pressures are small so that $\omega \simeq 0\Rightarrow $ $%
\chi (r)\simeq 1$. Thus, if there is enough pressure in the dark matter, $%
\chi (r)$ should have values away from unity.

\section{Application to our galaxy}

Zaritsky [45] collated the published older results (till 1998, see e.g.,
[49-53]) and demonstrated that they are all consistent with a galactic halo
that is nearly isothermal with a characteristic circular velocity
oscillating between $v_{\text{c}}\simeq 180$ to $220$ km sec$^{-1}$ at $15$
kpc. There are however more recent works on constraining the mass and extent
of the Milky Way's halo (see e.g., [54,55,56]). We shall use these data in
our computations below. Dehnen \textit{et al} [54] suggested a virial radius 
$R_{\text{vir}}\sim 200$ kpc, enclosing a virial mass $M_{\text{vir}}\sim
1.5\times 10^{12}M_{\odot }$. We adopt them as the halo radius $R_{\text{MW}%
} $ \ and mass $M_{\text{MW}}$ of our galaxy.

Xue \textit{et al} [55] observed that the Milky Way's circular velocity
curve at $\sim 60$ kpc gently falls from the adopted value of $v_{\text{c}%
}\simeq 220$ km sec$^{-1}$ at the Sun's location to $v_{\text{c}}\simeq 175$
km sec$^{-1}$ and implies $M$ ($<60$ kpc) $=(4.0\pm 0.7)\times
10^{11}M_{\odot }$. Deason \textit{et al} [56] infer that the mass within $%
150$ kpc probably lies in the range $(5-10)\times 10^{11}M_{\odot }$. The
measured fall in $v_{\text{c}}$ is not considered serious because the
implied mass ratio between the two extremes is only ($175/220$)$^{2}=0.63$.

Our strategy is to first find $\omega (r)$ from the Faber-Visser Eq.(21)
using the input of $v_{\text{c}}$ (that is, $\ell $) at some radius $r$.
Next, within the halo boundary $R_{\text{MW}}\sim 200$ kpc, we impose the
constraint $\omega (r<200$ kpc$)>0$, which means attractive dark matter
halo. At the boundary itself, we impose that $\omega (R_{\text{MW}})=0$,
thereby allowing for a change of sign in $\omega $ beyond the halo boundary.
We then analyze in detail the numerical limits on $\omega (r)$ using the
observed value of $\ell $ and different signs of the adjustable parameter $D$%
.

Following Xue \textit{et al} [55], we take $v_{\text{c}}(60$ kpc$)=175$ km
sec$^{-1}$, which means $\ell =2v_{\text{c}}^{2}/c_{0}^{2}=6.80\times
10^{-7} $. Then from Eq.(21), we get 
\begin{equation}
\omega (r)=\frac{6.43\times 10^{-14}+0.33Dr^{2}}{2.26\times 10^{-7}-Dr^{2}}%
>0,
\end{equation}%
which yields 
\begin{equation}
\omega (60\text{ kpc})=\frac{1.78\times 10^{-17}+0.33D}{6.30\times 10^{-11}-D%
}>0.
\end{equation}%
We now consider three cases of signs of $D$ and omit mentioning its
dimension in what follows.

\textbf{Case (1):} $D=0$. This value suggests itself. Then, from Eq.(26), we
have $\omega (r)=2.8\times 10^{-7}$ and $\chi (r)=1$ $\forall r$, which
imply that the perfect fluid approximates to dust dark matter. Because of
negligible pressure, as evidenced by the Faber-Visser function $\chi (r)=1$,
this case is more consistent with the Cold Dark Matter (CDM) paradigm for
galactic fluid [57]. We now use $\ell =6.80\times 10^{-7}$, $D=0$ in the
expressions for the masses $m_{\text{RC}}$, $m_{\text{Lens}}$, $M_{\text{pN}%
}(r)$ and find that all have nearly the same values\footnote{%
Note that the mass formulas in this paper are given in terms of distance $r$
kpc in relativistic units, but it is preferable to use the conventional and
direct unit of solar mass $M_{\odot }$. Therefore, we use the conversion $1$
kpc $=2.084\times 10^{16}M_{\odot }$ and re-express the masses in terms of $%
M_{\odot }$ in Sec.4.} within $r=60$ kpc, viz., $m_{\text{RC}}=m_{\text{Lens}%
}=M_{\text{N}}=M_{\text{pN}}=4.25\times 10^{11}M_{\odot }$ (Fig.1). The last
two equalities suggest that the model is Newtonian, that is, pressure
contribution is negligible [see Eqs.(23,24)]. Within the current level of
uncertainties in the values of observed mass, it is evident that our common
value is quite comparable with the value $M$ ($<60$ kpc) $=(4.0\pm
0.7)\times 10^{11}M_{\odot }$ inferred by Xue \textit{et al} [55]. Assuming
no further significant fall-off in the circular velocity from $v_{\text{c}%
}\simeq 175$ km sec$^{-1}$ (distinct from the radial velocity dispersion via
Jean's law), we find that at $r=150$ kpc, $m_{\text{RC}}=m_{\text{Lens}}=M_{%
\text{pN}}=1.0\times 10^{12}M_{\odot }$. This mass value is reasonably
consistent with the range $(5-10)\times 10^{11}M_{\odot }$ suggested by
Deason \textit{et al} [56].

Using $R_{\text{MW}}\sim 200$ kpc [54], and $\ell =6.80\times 10^{-7}$ [55], 
$D=0$, we find that, $m_{\text{RC}}=m_{\text{Lens}}=M_{\text{pN}}\sim
1.4\times 10^{12}M_{\odot }$ enclosed within the radius $R_{\text{MW}}$
(Fig.1). This mass value compares well with the virial mass $M_{\text{vir}%
}\sim 1.5\times 10^{12}M_{\odot }$ obtained by Dehnen \textit{et al }[54], $%
M_{\text{vir}}=1.0_{-0.2}^{+0.3}\times 10^{12}M_{\odot }$ obtained by Xue 
\textit{et al} [55], which is also comparable to\ the value $M_{\text{vir}}=$
$(1.26\pm 0.24)\times 10^{12}M_{\odot }$ obtained by McMillan [58] using a
Bayesian approach. Next, as is evident from Eqs.(10) and (24), generically, $%
M_{\text{total}}(r)\neq $ $M_{\text{pN}}(r)$ and it holds even in the case $%
D=0$, though the difference is negligible. Also, from Eq.(22), we find $4\pi
r^{2}\left[ p_{r}(r)-p_{t}(r)\right] \sim 10^{-14}$ $\forall r$, an
exceedingly small value consistent with the assumption of pressure isotropy
(Fig.2). However, nothing peculiar happens in $\omega $ at and beyond the
halo boundary $R_{\text{MW}}\sim 200$ kpc, because $\omega =$ $2.8\times
10^{-7}$ $\forall r$.

Another range suggested by Eq.(27) is $0<D<6.30\times 10^{-11}$, which also
leads to $\omega (r)>0$ $\ $for $r<60$ kpc from Eq.(26) but then there will
be an unphysical singularity in $\omega (r)$ appearing at $r_{\text{sing}}=%
\sqrt{2.26\times 10^{-7}/D}$, hence discarded here.

\textbf{Case (2):} $D<0$. We impose $\omega (R_{\text{MW}})=0$ ending the
extent of dark matter at the halo radius $R_{\text{MW}}\sim 200$ kpc. This
boundary condition yields, from Eq.(26), 
\begin{equation}
\omega (200\text{ kpc})=\frac{1.6\times 10^{-18}+0.33D}{5.67\times 10^{-12}-D%
}=0,
\end{equation}%
leading to a fixed value $D=-4.84\times 10^{-18}$. This value, when put back
in Eq.(26), leads to three different sectors: $\omega (r)>0$ for $r\in
\lbrack 0,200$ kpc$)$, $\omega (200$ kpc$)=0$ and $\omega (r>200$ kpc$)<0$
(Fig.3). This case has several interesting features and Fig.3 is the most
eloquent illustration of how the constraint Eq.(28) can eventually determine
the behavior of matter in all distance sectors.

First, the sector having $\omega (r)<0$ has positive energy density $\rho
(r)>0$ and negative pressure $p_{r}(r)+2p_{t}(r)<0$ (Fig.4), but the matter
is \textit{not} exotic as it still does not violate the Null Energy
Condition (NEC)\footnote{%
NEC is defined by $\rho +p_{r}\geq 0$ and $\rho +p_{t}\geq 0$. Matter that
violates these conditions is called "exotic".}(Fig.5). Interestingly, the
matter does not resemble either the cosmological phantom ($\omega <-1$) or
quintessence matter ($\omega <-1/3$) because, as is evident from Fig.3, $%
\omega (r)$ $\sim -10^{-7}$ at any finite $r>200$ kpc. However, at $%
r\rightarrow +\infty $, we find from Eq.(26) that $\omega (\infty
)\rightarrow (-1/3)+$, which therefore \textit{marginally} corresponds to
quintessence matter, and there appears no singularity in $\omega (r)$ at any
radius. Second, Fig.6 shows $\chi (r)\simeq 1$ for $r\in \lbrack 0,400$ kpc$%
] $, indicating that the pressure contribution is quite insignificant,
thereby once again supporting the CDM\ paradigm. Third, it can be easily
seen from Eq.(24) that the measure $M_{\text{pN}}(r)$ gives a comparable
galactic mass $\sim $ $1.4\times 10^{12}M_{\odot }$ enclosed within the
radius $R_{\text{MW}}$, while other mass estimates are also very close to
it. Fourth, from Eq.(22), it follows that $4\pi r^{2}\left[ p_{r}(r)-p_{t}(r)%
\right] \sim 10^{-14}$, a very minute difference as expected of perfect
fluid from the observational point of view. Finally, note that the value of $%
D$ actually determines the terminating radius where $\omega (r)=0$. For
example, for $D>-4.84\times 10^{-18}$, the radius can be arbitrarily shifted
away from $R_{\text{MW}}\sim 200$ kpc (for an illustration, see Fig.7). This
means that $D$ can be adjusted to the possibility of having a larger Milky
Way halo than considered here (viz., $200$ kpc).

In view of these merits,\ we can say that the range $-4.84\times 10^{-18}$ $%
\leq D\leq 0$, which in turn leads to $0\leq \omega (r)\leq 2.8\times
10^{-7} $ for the perfect fluid dark matter, can reasonably describe our
galactic halo. The simple physical requirement of an attractive halo thus
leads to a strong constraint indicating a dust-like dark matter $(\omega
\sim 0$ or $p\sim 0)$.

We wish to point out here that so far we focused only on the constraint from
the Milky Way, but there must be constraints from, for example, the Cosmic
Microwave Background (CMB) on deviations from $\omega =0$ for dark matter.
Using CMB, supernovae Ia and large scale structure data in a fluid dark
matter model, M\"{u}ller [59] found constraints on $\omega $ as follows: $%
-1.50\times 10^{-6}$ $<\omega <1.13\times 10^{-6}$ if the dark matter
produces no entropy and $-8.78\times 10^{-3}<\omega <1.86\times 10^{-3}$ if
the adiabatic sound speed vanishes, both at $3\sigma $ confidence level.
Clearly, we see that both the ranges in [58] concentrate around the value $%
\omega \sim 0$, which is in very good agreement with our result of a
dust-like halo. By observing effects of perturbations on CMB and matter
power spectra, Kumar and Lu [60] conclude that the current observational
data favor the CDM scenario with the cosmological constant type dark energy
at the present epoch. This is the most recent result on the CMB constraint. 

\textbf{Case (3):} $D>0$. Eq.(28) then suggests $0<D<5.67\times 10^{-12}$
giving $\omega (200$ kpc$)>0$. Let us take a concrete value, for example, $%
D=5\times 10^{-12}$ to see what that means. We find from Eq.(26) that $%
\omega (200$ kpc$)>0$ but $\omega \rightarrow \infty $ occurs at $r_{\text{%
sing}}=212$ kpc. Fig.8 shows the occurrence of cosmological quintessence
matter ($\omega <-1/3$) immediately beyond $212$ kpc. If $D>5\times 10^{-12}$%
, then $\omega (r)<0$ inside\textit{\ }the halo boundary, both contrary to
our assumption that $\omega (200$ kpc$)=0$. Fig.9 then shows that $M_{\text{%
pN}}(r)$ is larger than $M_{\text{N}}(r)$, meaning that there is \textit{%
substantial pressure} in the halo, as confirmed also by the $\chi $-factor
that shows values away from unity, signalling the presence of non-negligible
pressure as opposed to the CDM paradigm. All these features could make the
case truly intriguing if one is ready to live with a\ singularity in $\omega
(r)$. One might consider any other allowed value respecting $D<5.67\times
10^{-12}$, say $D=$ $2.83\times 10^{-12}$ (This particular value corresponds
to $\omega (200$ kpc$)=\frac{1}{3}$), then we find from Eq.(26) that the
singularity just shifts to a larger radius $r>200$ kpc, and quintessence
matter begins to appear from that radius onwards, as shown in Fig.10.
However, as we see the unphysical singularity can only be arbitrarily
shifted at will by choosing $D$ but not removed.

\section{Conclusions}

We revisited the perfect fluid dark matter model in the light of the
Faber-Visser post-Newtonian formalism that requires simultaneous measurement
of pseudo-mass profiles from rotation curve and gravitational lensing
observations. However, for the model considered here, no constraints from
lensing were used. The lensing mass together with other masses were computed
using recent data. The formalism provides information of the equation of
state of the galactic fluid, especially the pressure component in it. We saw
above how the variation of a crucial parameter $D$, that has dimension of
the cosmological constant (kpc)$^{-2}$, can lead to different scenarios. We
deduced the post-Newtonian version of isotropicity in Eq.(22) and computed
the equation of state parameter $\omega (r)$, the observables such as the
post-Newtonian mass $M_{\text{pN}}$, the rotation curve and lens
pseudo-masses from Eqs.(21,24,16,17) respectively in terms of the metric
functions\footnote{%
While the inferred pseudo-masses pertain to the same galaxy, they refer to
different radii, hence incomparable. The situation is likely to improve in
the future when observations with a higher resolution will be carried out
(see for details, [46]).
\par
{}}.

The computation of the above observables for our galaxy was done taking into
account the data on rotational velocity $\ell =6.80\times 10^{-7}$ from Xue 
\textit{et al} [55] and requiring an attractive halo ($\omega (r)\geq 0$) at
least within the halo radius $R_{\text{MW}}\sim 200$ kpc [54]. The choice of
the values of $D$ \ obtained from Eqs.(27,28), when used in Eq.(26), led to
the profiles of $\omega (r)$. Salient features of our analysis are
summarized below:

The case $D=0$ in Eq.(27) immediately led to $\omega (r)=2.8\times 10^{-7}$
and $\chi (r)=1$ $\forall r$, which imply that the perfect fluid
approximates to dust dark matter. The masses within $r=60$ kpc, viz., $m_{%
\text{RC}}=m_{\text{Lens}}=M_{\text{N}}=M_{\text{pN}}=4.25\times
10^{11}M_{\odot }$ are found to be in excellent agreement with the value $M$
($<60$ kpc) $=(4.0\pm 0.7)\times 10^{11}M_{\odot }$ inferred by Xue \textit{%
et al} [55]. The mass within $r=150$ kpc obtained by Deason \textit{et al}
[55] as well as the virial mass $M_{\text{vir}}$ from Dehnen \textit{et al}
[54] are also found to be quite comparable (Fig.1).

The case $D<0$ has a number of implications. The value $D=-4.84\times
10^{-18}$ corresponds to $\omega (200$ kpc$)=0$ ending dark matter halo.
Eq.(26) then leads to three different sectors: $\omega (r)>0$ for $r\in
\lbrack 0,200$ kpc$)$, $\omega (200$ kpc$)=0$ and $\omega (r>200$ kpc$)<0$
(Fig.3). The last sector has positive energy density $\rho (r)>0$ and
negative pressure $p_{r}(r)+2p_{t}(r)<0$ (Fig.4), but the matter is \textit{%
not} exotic as it still does not violate the Null Energy Condition
(NEC)(Fig.5). For $D>-4.84\times 10^{-18}$, the halo radius can be
arbitrarily shifted away from $200$ kpc (Fig.7), which means that $D$ can be
adjusted to the possibility of having a larger Milky Way halo than
considered here.

The case $D>0$ signals the presence of non-negligible pressure in the halo
as opposed to the CDM paradigm but also leads to a singularity in $\omega
(r) $ that can only be arbitrarily shifted at will by choosing $D$ but not
removed.

In view of the consistency with the recent galactic data and flexibility as
above, we suggest an overall range $-4.84\times 10^{-18}$ $\leq D\leq 0$,
which in turn leads to $0\leq \omega (r)\leq 2.8\times 10^{-7}$ for the
perfect fluid singularity-free equation of state of dark matter. As we see,
the values are concentrated around $D\sim 0$ leading to a strong constraint
of dust-like dark matter $(\omega \sim 0)$, which is supported also by CMB
constraints [59,60]. This is the main result of our paper.

We recall that an acceptable practical, working definition of a galactic
halo is still debatable [45]. One theoretically sound definition is that the
halo is the volume enclosing all of the mass that has already decoupled from
the Hubble flow. Another definition is the virial radius enclosing
gravitationally bound halo mass. All of them are problematic for practical
measurements. The only viable solution is to altogether avoid defining the
halo as a discrete entity. Instead, one should focus on the mass profile or
on the mass within a selected, fixed radius. However, all these arguments do
not rule out a future observation of a terminated discrete halo, even though
it might extend to hundreds of kpcs farther than the adopted $R_{\text{MW}%
}=200$ kpc. Within this ideology, the interval $0\leq \omega (r)\leq
2.8\times 10^{-7}$ allowing flexibility in the halo radius does make good
sense.

\section{Acknowledgment}

The authors wish to thank an anonymous referee for his/her insightful
comments that led to significant improvements over the initial version.

\section{References}

[1] F. Rahaman, K. K. Nandi, A. Bhadra, \ M. Kalam and K. Chakraborty, Phys.
Lett. B \textbf{694, }10 (2010).

[2] T. Harko and F.S. N. Lobo, Phys. Rev. D \textbf{83},124051 (2011).

[3] A.X. Gonz\'{a}lez-Morales and D. Nu\~{n}ez, J. Phys. Conf. Ser. \textbf{%
229}, 012041(2010).

[4] A. Avelino, Y. Leyva and L. A. Ure\~{n}a-L\'{o}pez, Phys. Rev. D\textbf{%
\ 88}, 123004 (2013).

[5] J. Oort, Bull. Astron. Inst. Netherl. \textbf{6}, 155 (1931).

[6] F. Zwicky, Astrophys. J.\textbf{\ 86}, 217 (1937).

[7] T. Matos and F. S. Guzm\'{a}n, Annalen d. Phys.\textbf{\ 9}, S1 (2000).

[8] F.S. Guzm\'{a}n and L.A. Ure\~{n}a-L\'{o}pez, Phys. Rev. D \textbf{68},
024023 (2003).

[9] T. Harko, F.S.N. Lobo, M.K. Mak and S.V. Sushkov, Mod. Phys. Lett. A 
\textbf{29}, 1450049 (2014).

[10] R. Izmailov, A.A. Potapov, A.I. Filippov, M. Ghosh and K.K. Nandi, Mod.
Phys. Lett. A \textbf{30},1550056 (2015).

[11] A.A. Potapov, R. Izmailov, O. Mikolaychuk, N. Mikolaychuk, M. Ghosh and
K.K. Nandi, JCAP 07(2015) 018.

[12] S. Nojiri and S.D. Odintsov, Phys. Rept.\textbf{\ 505}, 59 (2011).

[13] R.B. Metcalf and J. Silk, Phys. Rev. Lett. \textbf{98}, 071302 (2007).

[14] S. Bharadwaj and S. Kar, Phys. Rev. D \textbf{68}, 023516 (2003).

[15] M. Colpi, S.L. Shapiro and I. Wasserman, Phys. Rev. Lett. \textbf{57},
2485 (1986).

[16] G. Efstathiou, W.J. Sutherland and S.J. Maddox, Nature (London)\textbf{%
\ 348}, 705 (1990).

[17] T. Matos, F.S. Guzm\'{a}n and D. Nu\~{n}ez, Phys. Rev. D \textbf{62},
061301 (2000); T. Matos and L. A. Ure\~{n}a-Lopez, Phys. Lett. B \textbf{538}%
, 246 (2002).

[18] P.J.E. Peebles, Phys. Rev. D\textbf{\ 62}, 023502 (2000).

[19] U. Nucamendi, M. Salgado and D. Sudarsky, Phys. Rev. D \textbf{63}%
,125016 (2001).

[20] E.W. Mielke and F.E. Schunck, Phys. Rev. D \textbf{66}, 023503 (2002).

[21] J.E. Lidsey, T. Matos and L.A. Ure\~{n}a-Lopez, Phys. Rev. D \textbf{66}%
, 023514 (2002).

[22] A. Arbey, J. Lesgourgues and P. Salati, Phys. Rev. D \textbf{68},
023511 (2003).

[23] M.K. Mak and T. Harko, Phys. Rev. D \textbf{70}, 024010 (2004).

[24] G. Dvali, G. Gabadadze and M. Porrati. Phys. Lett. B \textbf{484,} 112
(2000); N. Arkani-Hamed, S. Dimopoulos and G. Dvali, Phys. Lett. B \textbf{%
429,} 263 (1998); I. Antoniadis, Phys. Lett. B \textbf{436,} 257 (1998); I.
Antoniadis, S. Dimopoulos and G. Dvali. Nucl. Phys. B \textbf{516,} 70
(1998).

[25] K.K. Nandi, A.I. Filippov, F. Rahaman, S. Ray, A.A. Usmani \textit{et
al., }Mon. Not. Roy. Astron. Soc. \textbf{399}, 2079 (2009).

[26] J.-c. Hwang and H. Noh, Phys. Lett. B \textbf{680}, 1 (2009).

[27] S.L. Dubovsky, P.G. Tinyakov and I.I. Tkachev, Phys. Rev. Lett. \textbf{%
94},181102 (2005).

[28] S. Dodelson and L.M. Widrow, Phys. Rev. Lett. \textbf{72},17 (1994).

[29] S. Dodelson, G. Gyuk and M.S. Turner, Phys. Rev. Lett. \textbf{72},
3754 (1994).

[30] P.D. Mannheim, Prog. Part. Nucl. Phys. \textbf{56}, 340 (2006).

[31] P.D. Mannheim and J.G. O'Brien, Phys. Rev. Lett. \textbf{106}, 121101
(2011).

[32] K.K. Nandi and A. Bhadra, Phys. Rev. Lett. \textbf{109}, 079001 (2012).

[33] M. Milgrom, Astrophys. J. \textbf{270}, 365 (1983); \textit{ibid.} 
\textbf{270}, 371 (1983);\textit{\ ibid.} \textbf{270}, 384 (1983).

[34] R.H. Sanders, Astrophys. J. \textbf{473},177 (1996).

[35] R.A. Swaters, R.H. Sanders and S.S. McGaugh, Astrophys. J. \textbf{718}%
, 380 (2010).

[36] J.W. Moffat, JCAP 03 (2006) 004.

[37] J.R. Brownstein and J.W. Moffat, Astrophys. J. \textbf{636}, 721 (2006).

[38] S. Capozziello, V.F. Cardone and A. Troisi, Mon. Not. Roy. Astron. Soc. 
\textbf{375}, 1423 (2007).

[39] A. Burkert, Astrophys. J.\textbf{\ 447}, L25 (1995).

[40] P. Salucci and A. Burkert, Astrophys. J. \textbf{53}7, L9 (2000).

[41] J.F. Navarro, C.S. Frenk and S.D.M. White, Astrophys. J. \textbf{490},
49 (1997).

[42] G. Jungman, M. Kamionkowski and K. Griest, Phys. Rept. \textbf{267},195
(1996).

[43] L.E. Strigari, Phys. Rept.\textbf{\ 531}, 1 (2013)

[44] G. Bertone, D. Hooper and J. Silk, Phys. Rept. \textbf{405}, 279 (2005).

[45] D. Zaritsky,\textit{\ Invited Review for The Third Stromlo Symposium:
The Galactic Halo} (1998) [ arXiv:astro-ph/9810069].

[46] T. Faber and M. Visser, Mon. Not. Roy. Astron. Soc. \textbf{372},136
(2006).

[47] K. Lake, Phys. Rev. Lett. \textbf{92}, 051101 (2004).

[48] T. Faber [arXiv:gr-qc/0607029].

[49] J. Einasto and D. Lynden-Bell, Mon. Not. Roy. Astron. Soc. \textbf{199}%
, 67 (1982).

[50] S. Raychaudhury and D. Lynden-Bell, Mon. Not. Roy. Astron. Soc. \textbf{%
240}, 195 (1989).

[51] P.J.E. Peebles, Astrophys. J. \textbf{449}, 52 (1995).

[52] E.J. Shaya, P.J.E. Peebles and R.B. Tully, Astrophys. J. \textbf{454},
15 (1995).

[53] M. Fich and S. Tremaine, Ann. Rev. Astron. Astrophys. \textbf{29}, 409
(1991).

[54] W. Dehnen, D.E. McLaughlin and J. Sachania, Mon. Not. Roy. Astron. Soc. 
\textbf{369}, 1688 (2006).

[55] X.-X. Xue \textit{et al}, Astrophys. J. \textbf{684}, 1143 (2008).

[56] A.J. Deason \textit{et al}, Mon. Not. Roy. Astron. Soc. \textbf{425},
2840 (2012).

[57] S. Dodelson, E. Gates and M.S. Turner, Science \textbf{274}, 69 (1996).

[58] P.J. McMillan, Mon. Not. Roy. Astron. Soc.\textbf{\ 414}, 2446 (2011).

[59] C.M. M\"{u}ller, Phys. Rev. D \textbf{71}, 047302 (2005).

[60] S. Kumar and L. Xu , Phys. Lett. B \textbf{737}, 244 (2014).

\begin{center}
--------------------------------------------------------------------------
\end{center}

\end{document}